\documentclass[preprint]{aastex}

\shorttitle{Asteroseismic verification of the KIC}
\shortauthors{Verner et al.}

\begin{document}

\title{Verification of the Kepler Input Catalog from asteroseismology of solar-type stars}

\author{
  G.~A.~Verner\altaffilmark{1,2},
  W.~J.~Chaplin\altaffilmark{1},
  S.~Basu\altaffilmark{3},
  T.~M.~Brown\altaffilmark{4},
  S.~Hekker\altaffilmark{5,1},
  D.~Huber\altaffilmark{6},
  C.~Karoff\altaffilmark{7},
  S.~Mathur\altaffilmark{8},
  T.~S.~Metcalfe\altaffilmark{8},
  B.~Mosser\altaffilmark{9},
  P-O.~Quirion\altaffilmark{10},
  T.~Appourchaux\altaffilmark{11},
  T.~R.~Bedding\altaffilmark{6},
  H.~Bruntt\altaffilmark{7},
  T.~L.~Campante\altaffilmark{12},
  Y.~Elsworth\altaffilmark{1},
  R.~A.~Garc\'ia\altaffilmark{13},
  R.~Handberg\altaffilmark{7},
  C.~R\'egulo\altaffilmark{14,15},
  I.~W.~Roxburgh\altaffilmark{2},
  D.~Stello\altaffilmark{6},
  J.~Christensen-Dalsgaard\altaffilmark{7},
  R.~L.~Gilliland\altaffilmark{16},
  S.~D.~Kawaler\altaffilmark{17},
  H.~Kjeldsen\altaffilmark{7},
  C.~Allen\altaffilmark{18},
  B.~D.~Clarke\altaffilmark{19},
  F.~R.~Girouard\altaffilmark{18}
}

\altaffiltext{1}{School of Physics and Astronomy, University of Birmingham, Edgbaston, Birmingham, B15 2TT, UK}
\altaffiltext{2}{Astronomy Unit, Queen Mary, University of London, Mile End Road, London, E1 4NS, UK}
\altaffiltext{3}{Department of Astronomy, Yale University, P.O. Box 208101, New Haven, CT 06520-8101, USA}
\altaffiltext{4}{Las Cumbres Observatory Global Telescope, Goleta, CA 93117, USA}
\altaffiltext{5}{Astronomical Institute, `Anton Pannekoek', University of Amsterdam, PO Box 94249, 1090 GE Amsterdam, The Netherlands}
\altaffiltext{6}{Sydney Institute for Astronomy (SIfA), School of Physics, University of Sydney, NSW 2006, Australia}
\altaffiltext{7}{Department of Physics and Astronomy, Aarhus University, DK-8000 Aarhus C, Denmark}
\altaffiltext{8}{High Altitude Observatory and Scientific Computing Division, National Center for Atmospheric Research, Boulder, Colorado 80307, USA}
\altaffiltext{9}{LESIA, CNRS, Universit\'e Pierre et Marie Curie, Universit\'e Denis Diderot, Observatoire de Paris, 92195 Meudon Cedex, France}
\altaffiltext{10}{Canadian Space Agency, 6767 Boulevard de l'A\'eroport, Saint-Hubert, QC, J3Y 8Y9, Canada}
\altaffiltext{11}{Institut d'Astrophysique Spatiale, Universit\'e Paris XI -- CNRS (UMR8617), Batiment 121, 91405 Orsay Cedex, France}
\altaffiltext{12}{Centro de Astrof\'isica, Universidade do Porto, Rua das Estrelas, 4150-762 Porto, Portugal}
\altaffiltext{13}{Laboratoire AIM, CEA/DSM -- CNRS -- Universit\'e Paris Diderot -- IRFU/SAp, 91191 Gif-sur-Yvette Cedex, France}
\altaffiltext{14}{Departamento de Astrof\'{\i}sica, Universidad de La Laguna, E-38206 La Laguna, Tenerife, Spain}
\altaffiltext{15}{Instituto de Astrof\'{\i}sica de Canarias, E-38200 La Laguna, Tenerife, Spain}
\altaffiltext{16}{Space Telescope Science Institute, Baltimore, MD 21218, USA}
\altaffiltext{17}{Department of Physics and Astronomy, Iowa State University, Ames, IA 50011, USA}
\altaffiltext{18}{Orbital Sciences Corporation/NASA Ames Research Center, Moffett Field, CA 94035, USA}
\altaffiltext{19}{SETI Institute/NASA Ames Research Center, Moffett Field, CA 94035, USA}

\begin{abstract}
We calculate precise stellar radii and surface gravities from the asteroseismic analysis of over 500 solar-type pulsating stars observed by the \emph{Kepler} space telescope.  These physical stellar properties are compared with those given in the \emph{Kepler Input Catalog} (KIC), determined from ground-based multi-colour photometry.  For the stars in our sample, we find general agreement but we detect an average overestimation bias of 0.23\,dex in the KIC determination of $\log(g)$ for stars with $\log(g)_\mathrm{KIC}>4.0$\,dex, and a resultant underestimation bias of up to 50\,\% in the KIC radii estimates for stars with $R_\mathrm{KIC}<2$\,$R_\odot$.  Part of the difference may arise from selection bias in the asteroseismic sample; nevertheless, this result implies there may be fewer stars characterised in the KIC with $R\sim 1$\,$R_\odot$ than is suggested by the physical properties in the KIC.  Furthermore, if the radius estimates are taken from the KIC for these affected stars and then used to calculate the size of transiting planets, a similar underestimation bias may be applied to the planetary radii.
\end{abstract}


\section{Introduction}

The primary task of the NASA \emph{Kepler Mission} is to detect the periodic dips in brightness caused by extrasolar planets transiting their host stars \citep{borucki2010}.  To accomplish this, the \emph{Kepler} space telescope simultaneously monitors the brightness of up to 150,000 stars.  While most of these stars are observed with a mean `long' cadence of 29.4 minutes, a subset of up to 512 stars at any one time may be observed with a mean `short' cadence of 58.85 seconds.  These high-precision, short-cadence observations make the data suitable for asteroseismic studies of solar-type stars \citep{gilliland2010,garcia2011}, which can precisely determine the stellar structure from their resonant modes of oscillation.

During the complete \emph{Kepler Mission}, approximately 160,000 stars will be surveyed in a field covering over 100 square degrees.  The Stellar Classification Project (SCP) was initiated in the pre-launch phase of the \emph{Kepler Mission} to carry out ground-based photometric classification of the targets in the \emph{Kepler} field and to construct the \emph{Kepler Input Catalog} (KIC) \citep{brown2011}.  The primary task of the SCP was to accurately distinguish giants from dwarfs in order to select solar-type stars which may have detectable transits of planets orbiting in the habitable zone.  The KIC also provided estimates of $T_\mathrm{eff}$, $\log(g)$, [Fe/H] and $E(B-V)$ from the analysis of multi-colour photometry, and derived properties (radius, mass and luminosity) from a simplified mass-radius relationship.  Since most of the stars in the \emph{Kepler} field-of-view had not been studied prior to the SCP, the classifications given in the KIC are often the only physical data available.

In order to characterise the structure of a super-Earth, the planetary radius should be known to a precision of $\sim5$\,\% and the mass to within 10\,\% \citep{valencia2007}.  To confirm the detection of an Earth-sized exoplanet based on its size requires the planetary radius to be similarly well-defined.  As the transit signature only gives the planetary radius relative to that of the host star, it is essential to obtain an estimate of the stellar radius to a similar precision in order to fully characterise an exoplanetary system.

The survey phase of the \emph{Kepler} asteroseismology program took place during the first 10 months of science operations.  Over this time more than 2600 stars were observed at short cadence, most for a duration of approximately one month.  A small number were observed for the complete survey phase, giving higher frequency resolution in their acoustic power spectra.  The asteroseismic properties of some of these stars have been analysed in detail, identifying individual p-mode frequencies that have been used to model their structures \citep{chaplin2010,metcalfe2010,campante2011,mathur2011}.  The average asteroseismic parameters have been measured in a large ensemble of stars observed during the survey and have been used to determine physical properties and to test synthetic stellar populations \citep{chaplin2011}.

In this paper we calculate precise radii and surface gravities from the average asteroseismic parameters of 514 solar-type stars observed during the survey.  These stars cover a range in apparent magnitude from 7.0 to 12.0 and a range in $T_\mathrm{eff}$ from 4400\,K to 6900\,K.  We use these asteroseismic structural parameters to verify the accuracy and precision of the physical properties given in the KIC.

\section{Kepler Input Catalog}

There is insufficient telemetry bandwidth to monitor all of the stars in the \emph{Kepler} field-of-view so targets are selected in order to maximise the probability of detecting a planetary transit around a solar-type star.  Of particular interest is the detection of an Earth-sized planet in the habitable zone of a Sun-like star.  Therefore, the primary task of the KIC is to correctly distinguish giants from dwarfs.

The aims and methods used for stellar classification in the KIC are described in detail by \citet{brown2011}.  Here we give a brief summary.  The physical data were derived from ground-based photometric observations in five bands ($g,r,i,z,D51$).  The stellar values of $T_\mathrm{eff}$, $\log(g)$, [Fe/H] and $E(B-V)$ were determined using Bayesian posterior probability maximisation to match the observed colours to stellar atmosphere models.  The values of $\log(g)$ derived from this method were then used directly to identify giants, defined as having $\log(g)<3.6$\,dex.

The KIC also gives estimates of radii, which were derived from $\log(g)$ assuming a mass-radius relationship that depends only on $T_\mathrm{eff}$.  As stated explicitly by \citet{brown2011}, this is true only in a statistical sense and takes no account of composition or evolutionary state.  The accuracy of the primary stellar properties in the KIC is stated to be 0.4\,dex for $\log(g)$ and $\pm$\,200\,K for $T_\mathrm{eff}$, with the ability to distinguish main-sequence stars from giants estimated to be reliable to 98\,\% confidence for stars with $T_\mathrm{eff}\leq5400$\,K \citep{koch2010,brown2011}.

\section{Asteroseismic data}

The frequency of maximum oscillation power, $\nu_\mathrm{max}$, and the average large frequency separation, $\Delta\nu$, can be obtained from the one-month \emph{Kepler} light curves using a number of different techniques \citep{mosser2009,huber2009,campante2010,hekker2010,mathur2010,verner2011}.  These characteristic asteroseismic parameters are sensitive to the globally-averaged stellar structure.  To obtain estimates of $\log(g)$ and $R$ from $\nu_\mathrm{max}$ and $\Delta\nu$, we used the Yale-Birmingham (YB) pipeline grid-based technique \citep{basu2010,gai2011}.  We tested the results obtained from the YB pipeline against those of similar methods \citep{stello2009a,quirion2010} and found them to be consistent.

The grid-based method determines the characteristics of stars by finding the maximum of the likelihood function of the set of input parameters 
$\{\nu_\mathrm{max},\Delta\nu,T_\mathrm{eff},[\mathrm{Fe}/\mathrm{H}]\}$ calculated with respect to a grid of stellar evolution models.  The uncertainties on the derived $\log(g)$ and $R$ were determined from the observational uncertainties on the input parameters and the priors that had been used to construct the stellar models.

It should be noted that the seismic parameters calculated for the stellar models rely on the approximate scaling relations:
\begin{equation}
  \nu_\mathrm{max} \approx \frac{M/M_\odot (T_\mathrm{eff}/T_{\mathrm{eff},\odot})^{3.5}}{L/L_\odot} \nu_{\mathrm{max},\odot} ,
  \label{eq:numax}
\end{equation}
\begin{equation}
  \Delta\nu \approx \frac{(M/M_\odot)^{0.5} (T_\mathrm{eff}/T_{\mathrm{eff},\odot})^3}{(L/L_\odot)^{0.75}} \Delta\nu_\odot .
  \label{eq:dnu}
\end{equation}

Equation (\ref{eq:dnu}) assumes that the average large frequency separation scales as the square root of the mean stellar density.  This relation has previously been tested using a wide range of stellar models \citep{ulrich1986,stello2009b,basu2010,white2011}.  \citet{white2011} tested the validity of Equation (\ref{eq:dnu}) using models with a broad range of masses and evolutionary states and found a deviation from the scaling relation that is predominantly a function of $T_\mathrm{eff}$.  According to \citet{white2011}, any systematic error introduced as a result of the deviation from Equation (\ref{eq:dnu}) is largest for low-mass stars with $T_\mathrm{eff} \approx 5600$\,K, where the value of $\Delta\nu$ calculated for the stellar models may be underestimated by up to $2.3$\,\%, and smallest for stars with $T_\mathrm{eff} \approx 4900$\,K and $T_\mathrm{eff} \approx 6400$\,K, where any deviation from Equation (\ref{eq:dnu}) is negligible.  For main-sequence stars with masses above $\sim1.2$\,$M_\odot$, which includes the majority of the stars in our sample, \citet{white2011} state that it is best not to include their correction to the scaling law.  As the stars in our sample cover a range of metallicities, masses and effective temperatures where any correction to the scaling relation is small, there should be no average systematic bias introduced as a result of assuming Equation (\ref{eq:dnu}).

Equation (\ref{eq:numax}) relies on the assumption that the frequency of maximum power should scale with the acoustic cut-off frequency \citep{brown1991,kjeldsen1995,belkacem2011}.  Unlike Equation (\ref{eq:dnu}), this has not been tested thoroughly with stellar models because there is less confidence in the theoretical predictions of the excitation and damping rates.  The $\nu_\mathrm{max}$ scaling has been tested empirically by \citet{chaplin2011} who concluded that it should be reliable to a few percent and that the net bias on stellar properties as a result of assuming both scaling laws is no larger than the quoted uncertainties obtained from the grid method.  The asteroseismic scaling laws are discussed in more detail and tested using \emph{Kepler} observations of main-sequence and giant stars in \citet{huber2011}.

The KIC determinations of $\log(g)$, $T_\mathrm{eff}$ and [Fe/H] were performed simultaneously, therefore to obtain a self-consistent comparison we used only the values of $T_\mathrm{eff}$ and [Fe/H] given in the KIC as input to the seismic analysis.  In the case of [Fe/H], the values given in the KIC are known to be imprecise.  We therefore adopted a very large uncertainty of 0.5\,dex in [Fe/H] so that the seismically determined values of $\log(g)$ and $R$ were not significantly sensitive to metallicity.  The large uncertainty in [Fe/H] is reflected in the derived uncertainties in $\log(g)$ and $R$.

It has been suggested that the values of $T_\mathrm{eff}$ given in the KIC are systematically too low for solar-type stars.  Pinsonneault et al. (in preparation) have used an alternative analysis of the raw photometry in the KIC, based on the method described in \citet{an2009}, and found $T_\mathrm{eff}$ that are on average $\sim5$\,\% higher than those given in the KIC.  A spectroscopic analysis of a sample of $\sim100$ stars has been performed by Bruntt et al. (in preparation) who found $T_\mathrm{eff}$ values that are on average $\sim3$\,\% higher than those in the KIC.  However, as it was our intention to test the KIC values of $\log(g)$ and the radii derived from these, both of which depend on the KIC derived $T_\mathrm{eff}$, we used the KIC estimates of $T_\mathrm{eff}$ to calculate the seismic values for $\log(g)$ and $R$.

Using the KIC $T_\mathrm{eff}$ provided a uniform set of temperatures for the stars under study and ensured that any systematic bias was not introduced as a result of using temperatures from another catalog.  It should be noted that due to the weak dependence of $\log(g)$ on $T_\mathrm{eff}$, an increase of 5\,\% in $T_\mathrm{eff}$ corresponds to an increase of only 0.01\,dex in the seismically determined $\log(g)$, and a decrease of only 1.2\,\% in the seismically determined $R$.

To test the robustness of the seismically determined values of $\log(g)$ and $R$, we applied the grid-based method to the sets of results for $\nu_\mathrm{max}$ and $\Delta\nu$ determined by five of the methods described in \citet{verner2011} that provided the greatest coverage over all stars with detected oscillations.  The YB pipeline was run with three different underlying model grids to determine the stellar properties \citep{dotter2008,marigo2008,gai2011}, giving up to 15 values of $\log(g)$ and $R$ for each star.

The mean uncertainty in $\log(g)$ calculated using all methods and model grids was 0.025\,dex.  The mean standard deviation of the results from all methods and model grids was 0.007\,dex.  In the case of $R$, the mean relative uncertainty was 7.0\,\%, while the mean standard deviation was 2.7\,\%.  The seismically determined stellar properties were therefore found to be insensitive to the choice of model grid or method used to determine $\nu_\mathrm{max}$ and $\Delta\nu$ at the level of precision available.  Since the scatter in the results was significantly smaller than the stated uncertainties, we compared the mean values of $\log(g)$ and $R$ to those in the KIC.

\section{Verification of KIC stellar properties}

\begin{figure}
  \includegraphics[width=0.5\hsize]{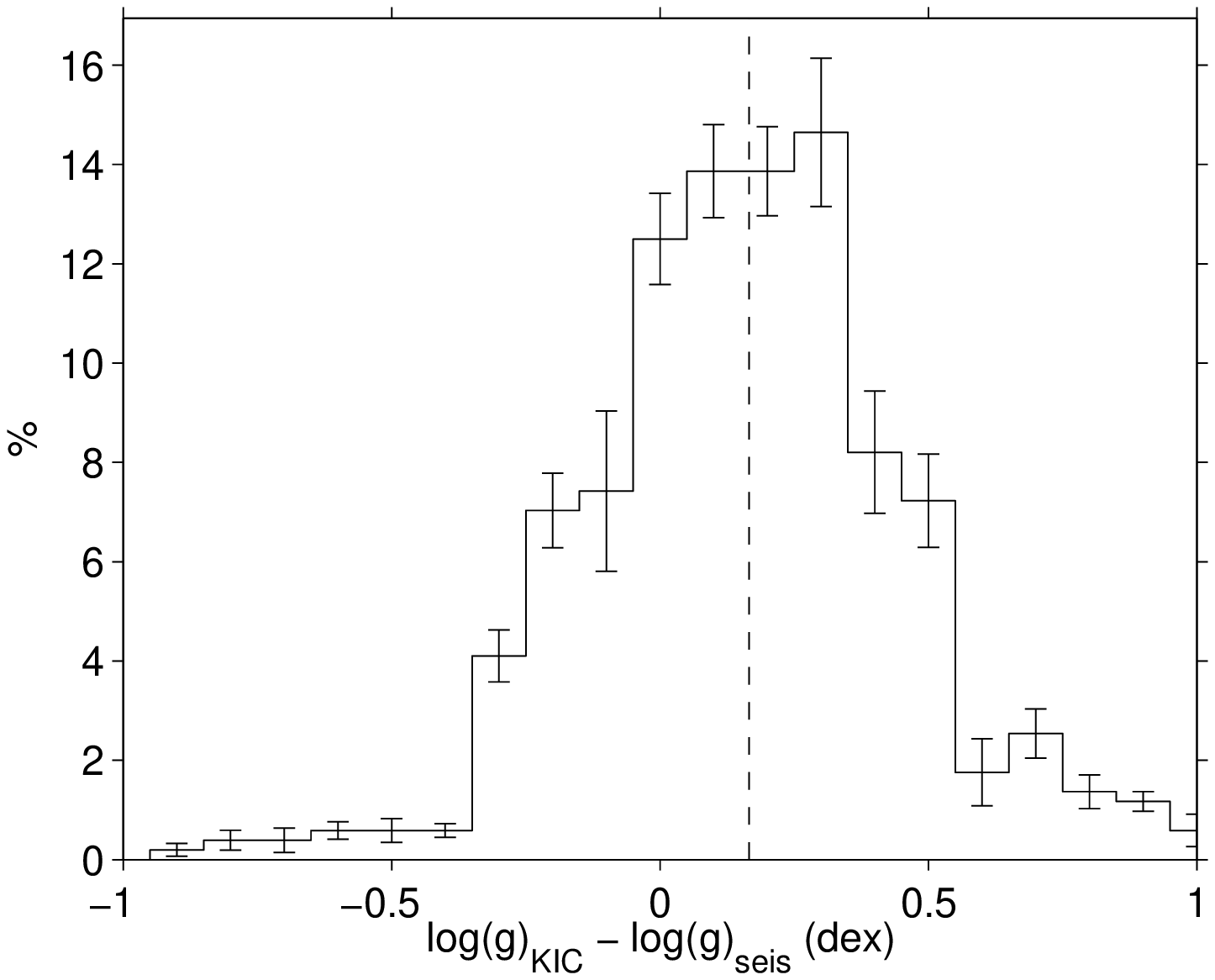}
  \medskip
  \includegraphics[width=0.5\hsize]{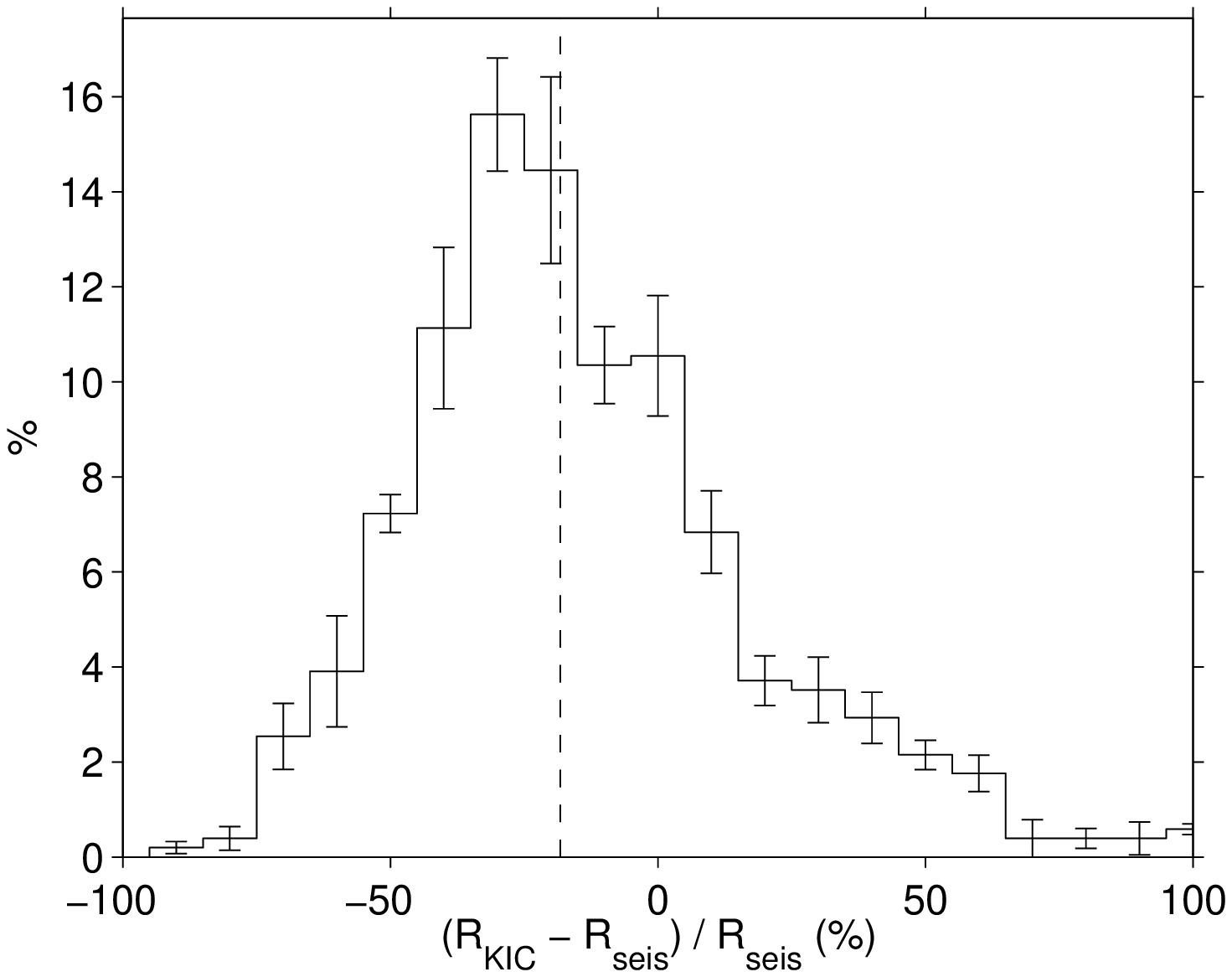}
  \caption{Distributions of the difference between KIC and seismic values of $\log(g)$ (left) and radius (right).  Error bars show the standard deviation in each range when determined using all possible combinations of model grid and input data.  Dashed lines show the median values of each distribution.}
  \label{fig:hist}
\end{figure}

The distributions of the differences between seismic and KIC values of $\log(g)$ and $R$ are shown in Fig.~\ref{fig:hist}.  As the seismic radii span a range from 0.7 to 6.4\,$R_\odot$, the relative differences rather than the absolute differences are used.  This necessarily causes a skewed distribution as the lower limit is fixed at $-100$\,\%.  The standard deviations of these distributions are 0.31\,dex ($\log(g)$) and 37\,\% ($R$).  The median values of the distributions indicate that there are significant biases in $\log(g)_\mathrm{KIC}$ ($+0.17$\,dex) and $R_\mathrm{KIC}$ ($-18$\,\%).

Since the radii given in the KIC were derived directly from $\log(g)$, an overestimation bias in $\log(g)$ causes an underestimation bias in $R$, since $M=gR^2$ (in solar units).  Through this relationship, for a star with $\log(g)=4.0$\,dex, an overestimation of $0.17$\,dex in $\log(g)$ results in an underestimation of $18$\,\% in $R$.

\subsection{\boldmath{$\log(g)$}}
\label{sect:logg}

\begin{figure}
  \includegraphics[width=0.5\hsize]{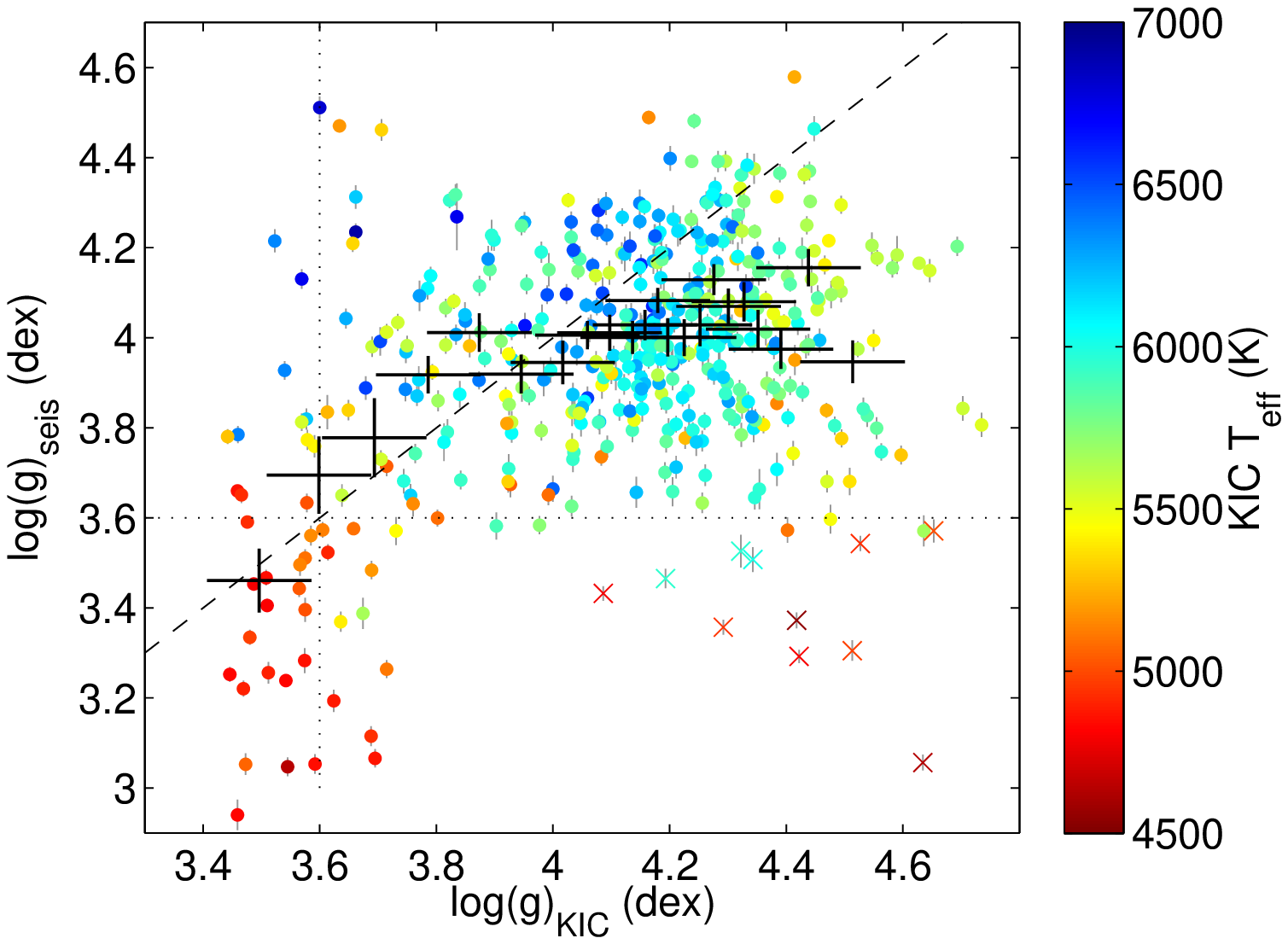}
  \medskip
  \includegraphics[width=0.5\hsize]{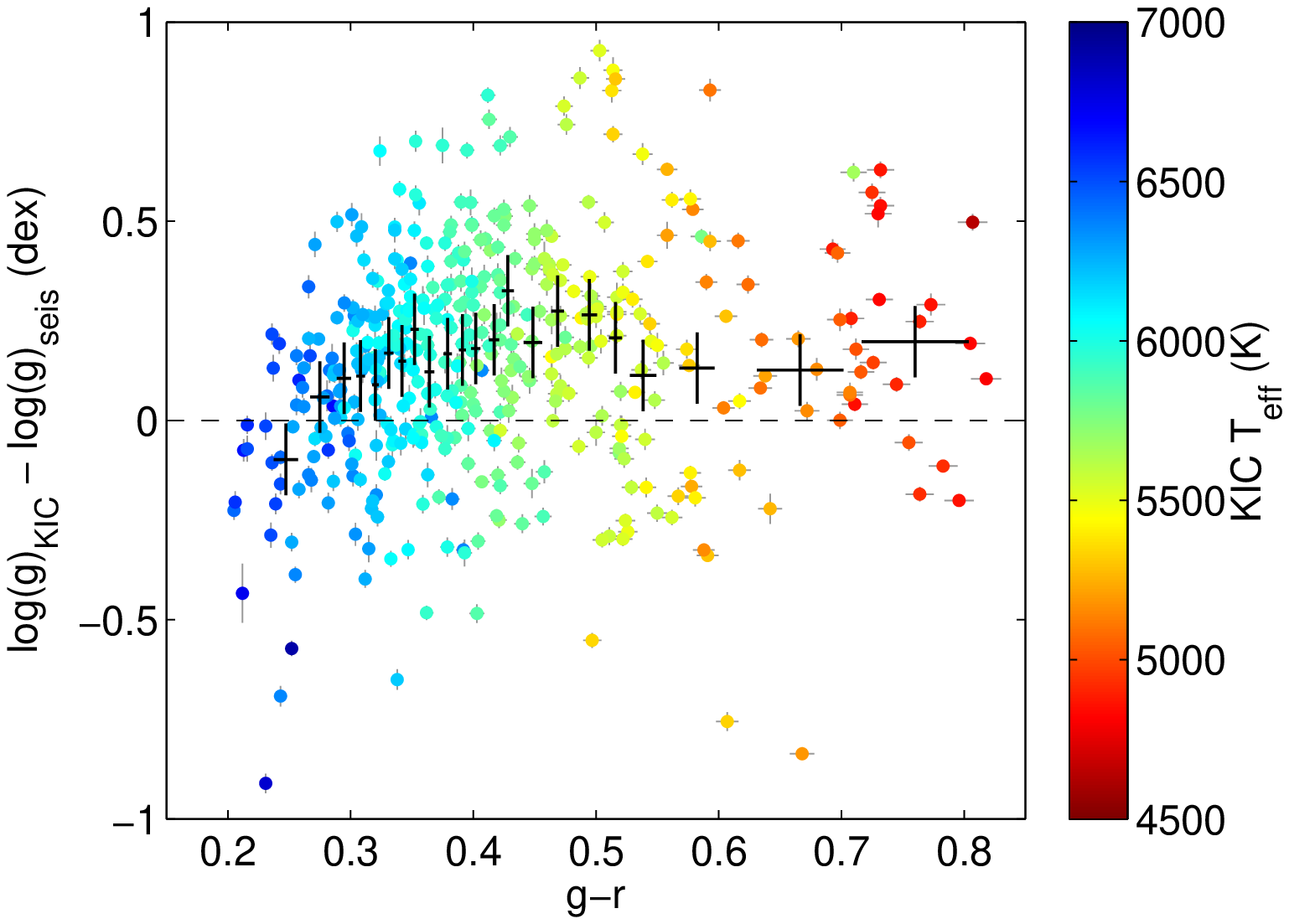}
  \caption{Comparison of KIC and seismic determination of $\log(g)$.  \textit{Left}: Dashed line shows line of equality, dotted lines show the cut-off point for giants, crosses show stars identified as giants by seismology but not by the KIC, error bars show 20-point averages (excluding misidentified giants).  \textit{Right}: $\log(g)$ difference as a function of the colour index $(g-r)$ taken from the KIC photometry.}
  \label{fig:logg}
\end{figure}

A comparison of seismic and KIC values of $\log(g)$ is shown in Fig.~\ref{fig:logg}.  Assuming an uncertainty of 0.4\,dex for $\log(g)_\mathrm{KIC}$, we find 11 stars (2\,\% of our sample) that would be identified as dwarfs using $\log(g)_\mathrm{KIC}$ but are classified as giants using asteroseismology (these stars are plotted as crosses in Fig.~\ref{fig:logg}).

We find that, on average, the KIC $\log(g)$ follows the seismic $\log(g)$ when $\log(g)_\mathrm{KIC}<4.0$\,dex.  There is no bias apparent in the $\log(g)$ differences for these stars and the $1\sigma$ agreement is at the 0.27\,dex level.  This agrees with the results in \citet{hekker2011} who showed that the values of $\log(g)$ in the KIC are consistent with those determined from the seismology of large cohort of red giants observed by \emph{Kepler}, with the majority having $2.3<\log(g)<2.7$\,dex.  The agreement becomes poorer for stars with $\log(g)_\mathrm{KIC}>4.0$\,dex, with the values in the KIC on average being systematically higher than the seismically determined $\log(g)$.  This behaviour was noted by \citet{brown2011}, who state that the KIC classifications tend to give $\log(g)$ too large for subgiants.

There are two observational biases that may contribute to this discrepancy.  Firstly, Malmquist bias increases the fraction of intrinsically brighter objects in the field, increasing the proportion of subgiants in the sample relative to their true number density.  Secondly, the amplitude of oscillations, and therefore their detectability, scales approximately with $L/M$, and therefore oscillations are preferentially detected for stars with lower $\log(g)$.  These two observational effects combine to give a relatively high fraction of subgiants in the seismic sample.  This is evident in the seismic $\log(g)$ distribution, which peaks around a value of 4.0\,dex.  The known shortfalls of the KIC lead to the overestimation of $\log(g)$ for these subgiants.  The average overestimation bias for stars with $\log(g)_\mathrm{KIC}>4.0$\,dex is 0.23\,dex with a $1\sigma$ scatter about this bias of 0.30\,dex.

Our findings confirm the warning given by \citet{brown2011} that users should be wary of $\log(g)_\mathrm{KIC}$ values for stars with photometric colours $(g-r)\leq0.65$.  Most of the stars in the seismic sample fall into this range.  The lower panel of Fig.~\ref{fig:logg} shows the difference between KIC and seismic $\log(g)$ as a function of $(g-r)$.  In the region where the KIC has shortfalls in the determination of $\log(g)$, the results from seismology can be used to complement the KIC results and perform an empirical correction.  We note that the absolute photometry in the KIC is very reliable, with an estimated error of just 1.5\,\% on the $(g-r)$ colour estimates.

\subsection{Radius}

The KIC provides estimates of stellar radii derived from $\log(g)$ and a simplified mass-radius relationship.  Asteroseismology can provide more precise estimates of the stellar radius \citep{jcd2010,metcalfe2010}, however most stars for which transits are observed will not have detectable oscillations.

\begin{figure}
  \includegraphics[width=0.5\hsize]{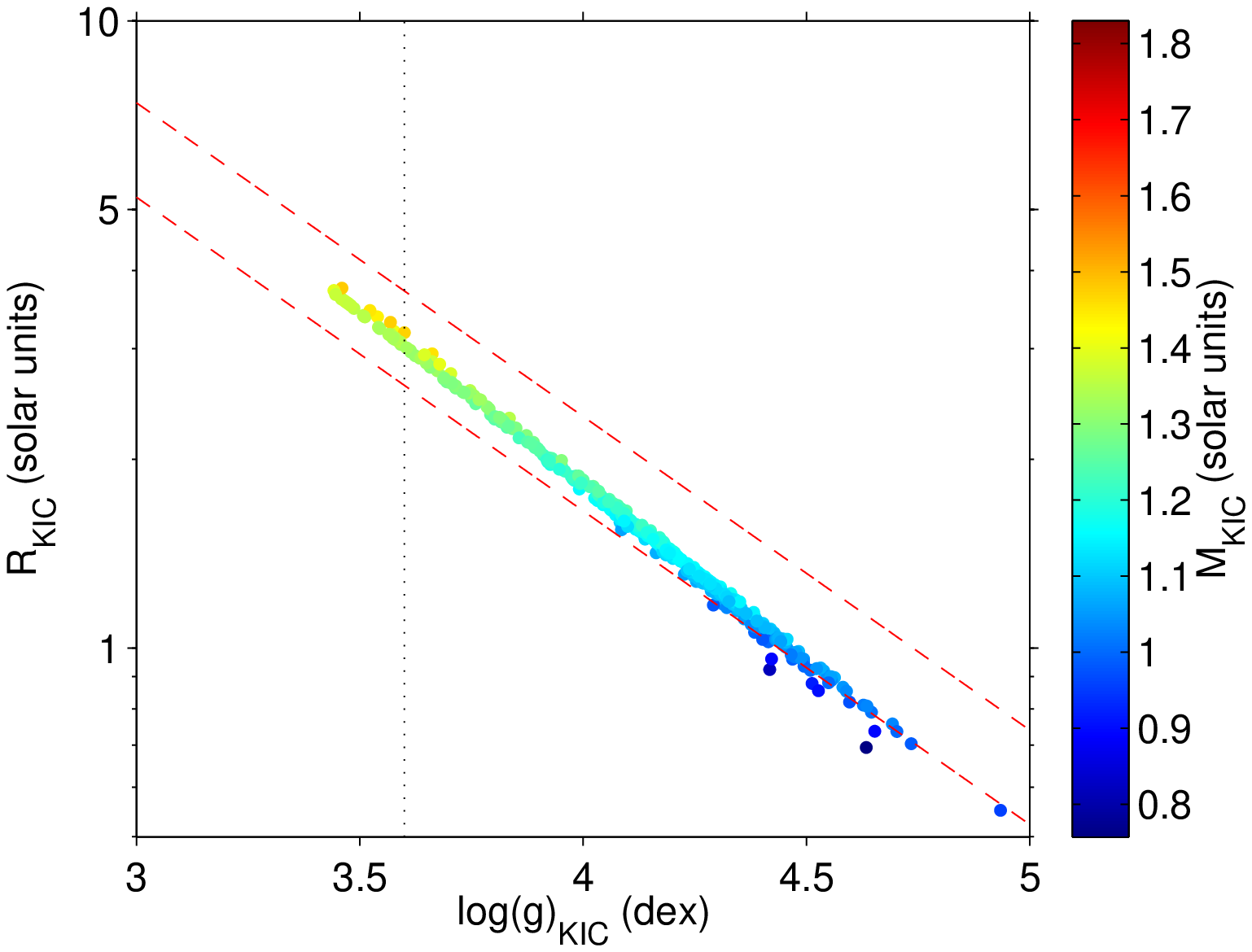}
  \medskip
  \includegraphics[width=0.5\hsize]{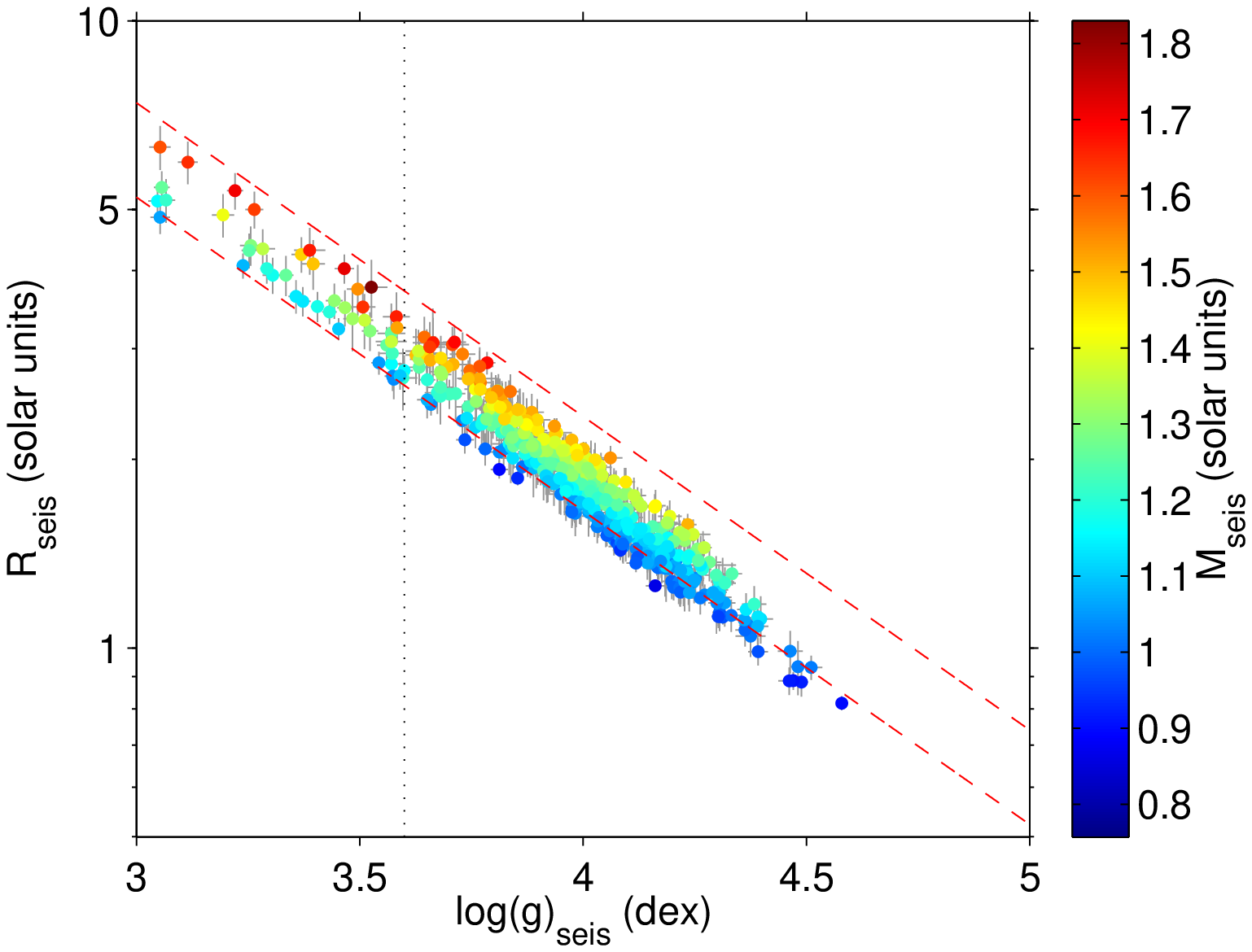}
  \caption{Radius--$\log(g)$ relationship using data from the KIC (left) and seismology (right).  Dashed lines show lines of constant mass for 1\,$M_{\odot}$ (bottom) and 2\,$M_{\odot}$ (top).  Dotted lines show the cut-off point separating giants from dwarfs.}
  \label{fig:massrel}
\end{figure}

The relationships between $\log(g)$, mass and radius are shown in Fig.~\ref{fig:massrel} for the KIC and seismic values.  It is clear that there is more variation in radius for a given $\log(g)$ in the seismic data than appears in the KIC, which is not surprising because the KIC relationship is assumed to be valid only in a statistical sense.  Through the KIC mass-radius relationship, the overestimation bias identified in $\log(g)_\mathrm{KIC}$ (Sect.~\ref{sect:logg}) appears as an underestimation bias in $R_\mathrm{KIC}$.

\begin{figure}
  \includegraphics[width=0.5\hsize]{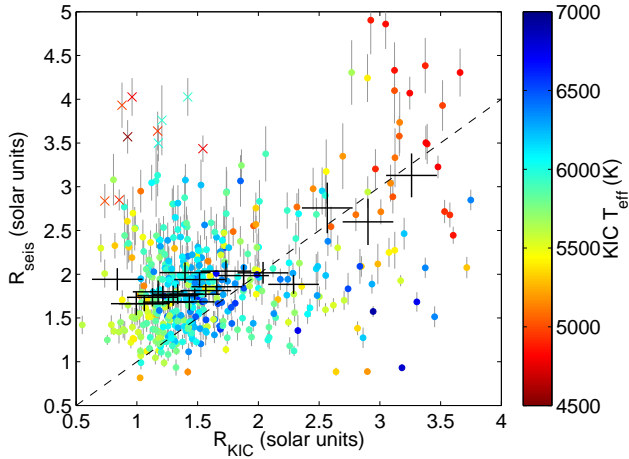}
  \caption{Comparison of KIC and seismic determination of stellar radius.  Dashed line shows line of equality, crosses show stars identified as giants by seismology but not by the KIC, error bars show 20-point averages.}
  \label{fig:radius}
\end{figure}

Fig.~\ref{fig:radius} compares the KIC radii with those determined from seismology.  An underestimation bias of up to 50\,\% is apparent for stars with $R_\mathrm{KIC}<2$\,$R_\odot$.  Using the KIC radii would imply that 20\,\% of the stars on which we have detected oscillations have radii within 0.2\,$R_\odot$ of the solar value.  However, the results from seismology prove that fewer than 3\,\% of the stars with detected oscillations are within this range.  It must be stressed that the same observational biases described in Sect.~\ref{sect:logg} apply equally to the results for stellar radii.  The cohort of stars that has been used in this comparison is subject to selection effects as this only includes the relatively small fraction of stars observed by \emph{Kepler} that have detected oscillations.  This is biased towards stars with higher $L/M$ and may not be representative of all stars in the KIC.

For stars with $R_\mathrm{KIC}>2$\,$R_\odot$, the standard deviation of the distribution of relative differences between KIC and seismic radii is 35\,\% and no average bias is apparent.  However, for stars with $R_\mathrm{KIC}>2.5$\,$R_\odot$, a dichotomy is present whereby cool stars ($T_\mathrm{eff}\lesssim5000$\,K) tend to fall above the agreement line and hot stars ($T_\mathrm{eff}\gtrsim6500$\,K) fall below it.  These stars correspond to the extremes of the $(g-r)$ plot shown in Fig.~{\ref{fig:logg}}.  For hot stars with $(g-r)<0.26$, the bias reverses sign and $\log(g)_\mathrm{KIC}$ is underestimated by up to 0.25\,dex, leading to an overestimation of $R_\mathrm{KIC}$ by up to 50\,\%.  Any empirical correction to the data in the KIC should therefore be done as a function of $(g-r)$ and not as a function of $\log(g)_\mathrm{KIC}$ or $R_\mathrm{KIC}$.

\section{Discussion}

Using precise values of $\log(g)$ determined from the asteroseismic analysis of a large ensemble of solar-like oscillators, we have found that in its primary role the KIC effectively distinguishes dwarfs from giants.  However, at the level of precision available to the asteroseismic parameters, we find an overestimation bias for stars with $\log(g)_\mathrm{KIC}>4.0$\,dex, although this bias may be overstated relative to the full catalog due to selection effects in the asteroseismic sample.

For stars with $\log(g)_\mathrm{KIC}<4.0$\,dex, on average we find an unbiased agreement between the seismic and KIC values, with a $1\sigma$ deviation of 0.27\,dex.  This precision is higher than was estimated in \citet{brown2011} (0.4\,dex).  However, due to the presence of a significant number of subgiants in our sample, and the known shortfalls of the KIC that result in an overestimation of $\log(g)$ for these stars, a representative uncertainty of 0.4\,dex for $\log(g)_\mathrm{KIC}$ is nevertheless probably appropriate.

The radii given in the KIC are affected by the overestimated values of $\log(g)$ and as a result may be underestimated by up to 50\,\% for stars with $R_\mathrm{KIC}<2$\,$R_\odot$.  For stars with $R_\mathrm{KIC}$ larger than this, the agreement is better with a relative $1\sigma$ deviation of 35\%.  This is important to take into account if the KIC radii are used to estimate the number of $\sim1$\,$R_\odot$ stars in our sample.  The results from asteroseismology show that a much smaller number of stars with detected oscillations have radii close to $1$\,$R_\odot$ when compared with the radii given in the KIC.  This bias may lead to an underestimation of the radii of exoplanets transiting the affected stars if the KIC radius is used to calculate the planetary radius \citep{borucki2011}, however the bias is strongest for subgiants and may be considered an upper limit for less evolved stars.

Although the KIC was not designed primarily to give values of $\log(g)$ and $R$ to high precision, we find that the catalog may be used to give an indication of these stellar properties once the limitations outlined above are taken into account.  Much better estimates of $\log(g)$ and $R$ are available from asteroseismology for stars with detected solar-like oscillations.

\section*{Acknowledgements}

GAV, WJC, YE and IWR acknowledge the support of the UK Science and Technology Facilities Council.  SH acknowledges support from the Netherlands Organisation for Scientific Research.  Funding for the \emph{Kepler Mission} is provided by NASA's Science Mission Directorate.  The authors wish to thank the entire \emph{Kepler} team, without whom these results would not be possible.  We also thank all funding councils and agencies that have supported the activities of KASC Working Group 1.

\clearpage

\end{document}